\documentclass[journal]{IEEEtran}
\ifCLASSINFOpdf
\else
\fi
\usepackage{cite}
\usepackage{amssymb}
\usepackage{amsmath}
\usepackage{graphicx}

\newtheorem{lemma}{Lemma}
\newtheorem{theorem}{Theorem}

\newtheorem{definition}{Definition}
\newtheorem{remark}{Remark}
\newtheorem{example}{Example}
\usepackage{epstopdf}
\usepackage{color}
\usepackage{cases}

\usepackage{subfigure}
\begin{document}
%
\title{Self-Triggered Scheduling for Boolean Control Networks}
%
%
%

\author{
Min Meng, Gaoxi Xiao, {\em Senior Member, IEEE} and Daizhan Cheng, {\em Fellow, IEEE}
\thanks{M. Meng (minmeng@ntu.edu.sg;mengminmath@gmail.com) and G. Xiao (egxxiao@ntu.edu.sg) are with the School of Electrical and Electronic Engineering, Nanyang Technological University, 50 Nanyang Avenue, Singapore 639798. D. Cheng (dcheng@iss.ac.cn
) is with Key Laboratory of Systems and Control, Academy of Mathematics and Systems Sciences, Chinese Academy of Sciences, Beijing 100190, P. R. China.}
\thanks{Corresponding author: Gaoxi Xiao}
\thanks{This work was partially supported by Ministry of Education, Singapore, under contract of MOE2016-T2-1-119.}
}

\maketitle

\begin{abstract}
It has been shown that self-triggered control has the ability to reduce computational loads and deal with the cases with constrained resources by properly setting up the rules for updating the system control when necessary. In this paper, self-triggered stabilization of Boolean control networks (BCNs), including deterministic BCNs, probabilistic BCNs and Markovian switching BCNs, is first investigated via semi-tensor product of matrices and Lyapunov theory of Boolean networks. The self-triggered mechanism with the aim to determine when the controller should be updated is provided by the decrease of the corresponding Lyapunov functions between two consecutive sampling times. Rigorous theoretical analysis is presented to prove that the designed self-triggered control strategy for BCNs is well defined and can make the controlled BCNs be stabilized at the equilibrium point.
\end{abstract}


%
\IEEEpeerreviewmaketitle

\section{Introduction}
Boolean networks have attracted considerable attention due to their wide applications in various fields such as gene regulatory networks \cite{kauffman1969metabolic}, smart home \cite{kabir2014mathematical} and game theory \cite{alexander2003random,cheng2014finite,zhang2018incomplete}, etc.
Extensive studies have been conducted on analysis and control problems of Boolean networks by semi-tensor product of matrices \cite{Chengd2007,Chengd2011} in the last decade, with different focuses on system stability, optimization, observability, controllability and so on. Readers may refer to \cite{cheng2015receding,fornasini2013observability,guo2017invariant,liang2017improved,liu2017function,lu2016pinning,toyoda2019mayer,meng2019controllability,li2018survey,zhong2019pinning,li2020perturbation} and the references therein for more details.

As is well known, there are always switching and uncertain phenomena in practical systems. For example, the bacteriophage $\lambda$ in genetic regulatory networks may possess different behaviors (lysis and lysogeny) under different external and internal environments. A series of molecular processes in genetic regulatory networks is always affected by some intrinsic fluctuations and extrinsic perturbations with stochastic factors. Therefore, probabilistic/Markovian switching Boolean networks may have advantages in modeling the rule-based properties and uncertainties. Stability and stabilization of probabilistic/Markovian switching Boolean networks have been investigated in \cite{li2014state,li2018set,meng2017stability,meng2018stability,huang2020stability}.

In most existing references on stabilization and controller design for Boolean control networks (BCNs), it is required that the states at all the discrete time should be accessible. However, caused by constrained resources such as a limited lifetime of battery-powered devices, the received data from sensors for designing controllers may be disrupted.  In addition, it was pointed out \cite{li2016event,yue2017event} that in the study of genetic regulatory networks, the feedback control based on the data at all the consecutive discrete time may lead to some undesirable results such as too frequent transmission of information between mRNA and protein, Zeno behavior, and so on, which may consume a large number of controller executions and computational costs. Consequently, it is necessary to develop control techniques depending on the measurable states at partial discrete time.

 Periodic sampling, which is a special case where the measurements are available periodically, has been applied to study state feedback stabilization for BCNs \cite{liu2016sampled,zhu2018sampled}. The work was then extended to non-periodical sampling \cite{liu2019sampled}, which is also prescheduled. Such sampling intervals can be regarded as exogenous signals which are deterministic regardless of whether the systems need attention. On the other hand, however, the sampling time is always unknown in advance in event-based cases, where the next sampling time at which the control is updated always hinges on the control itself and a state-dependent criterion in a way that the stability of the closed-loop system is not destroyed \cite{aaarzen1999simple,tabuada2007event,lunze2010state,lehmann2011event}. Related works on event-based control of Boolean networks can be found in \cite{li2018event_tc,li2018event,zhu2019stabilizing,yang2019event,tong2018robust,lu2019event}. In \cite{li2018event_tc}, the disturbance decoupling problem was studied by event-triggered control and the triggering condition as a rank condition of the network transition matrices. In \cite{li2018event}, the authors designed the triggering times based on the Hausdorff distance to study robust control of BCNs with disturbances. Subsequently, Zhu and Lin in \cite{zhu2019stabilizing} obtained an optimal event-triggered control strategy for stabilization of BCNs by constructing the weighted digraph and the hypergraph for the BCN and applying the shortest path algorithm to the hypergraph.
The idea of event-triggered control was also extended to study synchronization of drive-response BCNs \cite{yang2019event} and robust invariance of probabilistic BCNs \cite{tong2018robust}. Such results can indeed reduce the number of samples while still fulfilling the requests.

The event-triggered control, with all its advantages, has to depend on constant measurements to detect whether the triggering conditions are fulfilled. However, self-triggered sampling scheduling \cite{velasco2003self,wang2009self,mazo2010iss} has the advantage that the next sampling time $t_{k+1}$ can be determined in advance only based on the state and controller at the current sampling time $t_k$.  To our best knowledge, there are no references on self-triggered control for BCNs, which motives our study in this paper for improving the existing periodic/event-triggered sampling schemes for BCNs.

In this paper, for the first time to the best of our knowledge, we investigate self-triggered scheduling for BCNs based on Lyapunov functions for Boolean networks. Three kinds of BCNs, namely deterministic BCNs, probabilistic BCNs and Markovian switching BCNs, respectively, are considered. Lyapunov functions for deterministic and Markovian switching Boolean networks were, respectively,  proposed in \cite{wang2012definition} and \cite{meng2017stability}. However, there is no systematic analysis on Lyapunov stability for all the different classes of Boolean networks. In this paper, the definition and construction of the Lyapunov function for probabilistic Boolean networks are presented. The self-triggered conditions are designed hinging on the known stabilizing controllers and the decrease of the corresponding Lyapunov functions between two consecutive samplings. The self-triggered controllers improve the known ones when only partial state information is available. Note that Boolean networks are a kind of nonlinear networks with finite states, then the methods and results for BCNs are not trivial and not similar to those of the conventional discrete-time systems. We provide rigorous theoretic analysis to prove that the presented self-triggered update scheduling is well defined for (deterministic, probabilistic, Markovian switching) BCNs and can make the controlled BCNs stable.

In summary, the main contributions of this paper are twofold:
\begin{itemize}
\item[i)] The definition of Lyapunov function and Lyapunov stability theory for probabilistic Boolean networks are presented for the first time, which can be applied to easily constructed a Lyapunov function for probabilistic Boolean networks.
\item[ii)] A self-triggered scheduling for BCNs is proposed based on the decrease of the constructed Lyapunov functions between two consecutive sampling. Rigorous analyses are given to show the well-definedness of the designed self-triggered controllers and stabilization for three kinds of BCNs, namely deterministic BCNs, probabilistic BCNs and Markovian switching BCNs, respectively.
\end{itemize}

The remainder of this paper is organized as follows. Section \ref{section2} introduces some preliminary results about semi-tensor product of matrices. In Section \ref{section_Ly}, we introduce the Lyapunov stability theory for three kinds of Boolean networks. In Section \ref{section3}, self-triggered scheduling and theoretical analysis are presented for BCNs, probabilistic BCNs and Markovian switching BCNs, respectively.  Finally, a brief conclusion is given in Section \ref{section5}.

\emph{Notations.}  Let $\mathbb{R}^n$ and $\mathbb{R}^{m\times n}$ be the sets of $n$-dimensional column vectors and $m\times n$ real matrices, respectively. Set $\mathbb{B}:=\{0,1\}$. The symbol $\mathbb{B}^{n\times m}$ represents the set of $n\times m$ matrices with every element being in $\mathbb{B}$. The matrices in $\mathbb{B}^{n\times m}$ are called Boolean matrices. $\mathbb{B}^n:=\mathbb{B}^{n\times 1}$. $\delta_n^i$ represents the $i$th column of the identity matrix $I_n$, $i=1,2,\ldots,n$. Denote $\Delta_n:=\{\delta_n^i\mid i=1,2,\ldots,n\}$. A matrix $L\in\mathbb{R}^{n\times r}$ is called a logical matrix if every column of $L$ is in $\Delta_{n}$, and a logical matrix $L\in\mathbb{R}^{n\times r}$ can be written as $L=[\delta_n^{i_1},\delta_n^{i_2},\ldots,\delta_n^{i_r}]$ or $L=\delta_n[i_1,i_2,\ldots,i_r]$. Denote by ${\mathcal L}_{n\times r}$ the set of $n\times r$ logical matrices.
${\rm Col}_i(L)$ represents the $i$th column of $L$ and ${\rm Col}(L)$ is the set of columns of $L$. $W_{[m,n]}$ represents an $mn\times mn$ swap matrix defined in \cite{Chengd2007,Chengd2011}, i.e., $W_{[m,n]}=[I_n\otimes\delta_{m}^1~I_n\otimes\delta_{m}^2~\cdots~I_n\otimes\delta_{m}^m]$, where  $\otimes$ is the Kronecker product \cite{liu2008Hadamard}. ${\bf 1}_n$ (${\bf 0}_n$) is a column vector in $\mathbb{R}^n$ with all of its elements being $1$ ($0$). ${\rm diag}\{M_1,M_2,\ldots,M_n\}$ represents a diagonal matrix with the $i$th diagonal being $M_i$. The notation $A<(>,\leq,\geq) ~{\bf 0}$ for a matrix or a vector $A$  means that all the elements of $A$ are negative (positive, nonpositive, nonnegative). $\rho(A)$ is the spectral radius of matrix $A$. The symbols ${\bf Pr}\{\cdot\}$ and ${\bf E}\{\cdot\}$ represent the probability and expectation operators, respectively.

\section{Semi-tensor product}\label{section2}
In this section, some preliminaries about semi-tensor product of matrices  are introduced.
We first give the definition of the main mathematical tool, semi-tensor product of matrices, used in this paper.
\begin{definition}[\cite{Chengd2007,Chengd2011}]\label{defdef1}
The semi-tensor product of matrices $M\in\mathbb{R}^{a\times b}$ and $N\in\mathbb{R}^{c\times d}$, denoted by $M\ltimes N$, is defined as
\begin{align*}
M\ltimes N=(M\otimes I_{l/b})(N\otimes I_{l/c}),
\end{align*}
where $l$ is the least common multiple of $b$ and $c$.
\end{definition}

When the column dimension of $M$ is equal to the row dimension of $N$, i.e., $b=c$, the semi-tensor product of $M$ and $N$ is degenerated to the traditional matrix product, i.e., $M\ltimes N=MN$. Hence, the STP is a generalization of conventional matrix product. Moreover, this generalization keeps all major properties of traditional matrix product, such as distributive law, associative law and so on. In this paper, the symbol ``$\ltimes$'' is omitted if no confusion arises.
Further discussions on properties and applications of semi-tensor product can be referred to \cite{Chengd2007,Chengd2011}.

The essential step of using semi-tensor product of matrices to study Boolean networks is to define a bijective mapping from $\mathbb{B}$ to $\Delta_2$, i.e., $0\sim\delta_2^2$, $1\sim\delta_2^1$. Then we can get a bijection from $\mathbb{B}^n$ to $\Delta_{2^n}$, denoted by $\phi_n: \mathbb{B}^n\rightarrow\Delta_{2^n}$, which is defined as
\begin{align}\label{equ1}
\phi_n(X)=\left(\begin{array}{c}X_1\\ \bar{X}_1\end{array}\right)\ltimes\left(\begin{array}{c}X_2\\ \bar{X}_2\end{array}\right)\ltimes\cdots\ltimes\left(\begin{array}{c}X_n\\ \bar{X}_n\end{array}\right)\in\Delta_{2^n},
\end{align} where $X=(X_1,X_2,\ldots,X_n)^T\in\mathbb{B}^n$ and $\bar{X}_i=1-X_i$, $i=1,2,\ldots,n$.
Note that a Boolean function with $n$ variables is a mapping from $\mathbb{B}^n$ to $\mathbb{B}$. An important lemma for equivalently converting the original logical form of Boolean networks to an algebraic expression is presented as follows.
\begin{lemma}[\cite{Chengd2007,Chengd2011}]\label{lemma1}
For a Boolean function $\psi:\mathbb{B}^n\rightarrow\mathbb{B}$, there exists a unique matrix $M_\psi\in{\cal L}_{2\times 2^n}$, which is named as the structure matrix of $\psi$, such that
\begin{align}
\phi_1(\psi(X))=M_\psi\phi_n(X),
\end{align} where $\phi_1,\phi_n$ are defined in (\ref{equ1}).
\end{lemma}

\section{Lyapunov stability theory}\label{section_Ly}
This section will introduce the Lyapunov stability theory for three classes of Boolean networks, namely deterministic Boolean networks, probabilistic Boolean networks and Markovian switching Boolean networks. The Lyapunov function for probabilistic Boolean networks is defined for the first time, while those for the other two kinds of Boolean networks can be found in \cite{wang2012definition,li2017lyapunov,meng2017stability,meng20161}. Without loss of generality, it can be assumed that the equilibrium point of a (deterministic, probabilistic or Markovian switching) Boolean network is $\delta_{2^n}^{2^n}$. Otherwise, a coordinate transformation \cite{cheng2010realization} can be used to equivalently transfer any point $\delta_{2^n}^k$ to $\delta_{2^n}^{2^n}$. We start by emphasizing that the notations in Subsections \ref{sub2.2}, \ref{sub2.3} and \ref{sub2.4} are independent.
\subsection{Lyapunov function for Boolean networks}\label{sub2.2}
A Boolean network with $n$ nodes is given as
\begin{align}\label{equ3}
X(t+1)=f(X(t)),
\end{align} where $X(t)\in\mathbb{B}^n$ and $f:\mathbb{B}^n\rightarrow\mathbb{B}^n$ is a Boolean vector function. Based on the semi-tensor product in Definition \ref{defdef1} and Lemma \ref{lemma1}, the algebraic form of Boolean network (\ref{equ3}) can be equivalently rewritten as
\begin{align}\label{equ4}
x(t+1)=Fx(t),
\end{align} where $x(t)=\phi_n(X(t))\in\Delta_{2^n}$, and $F$ is in ${\cal L}_{2^n\times2^n}$, called the transition matrix of (\ref{equ3}). Partition $F$ as
$$F=\left[\begin{array}{cc}F_{11}&F_{12}\\F_{21}&F_{22}\end{array}\right],$$ where $F_{11}\in\mathbb{B}^{(2^n-1)\times(2^n-1)}$ and $F_{22}\in\mathbb{B}$.
Then one can verify that a necessary condition for stability at the equilibrium point $\delta_{2^n}^{2^n}$ of Boolean network (\ref{equ4}) is that $\delta_{2^n}^{2^n}$ is a fixed point of (\ref{equ4}), which is equivalent to $F\delta_{2^n}^{2^n}=\delta_{2^n}^{2^n}$, i.e., $F_{12}={\bf 0}_{2^n-1}$ and $F_{22}=1$.
Reviewing the Lyapunov theory proposed in \cite{wang2012definition,li2017lyapunov,meng20161}, Boolean network (\ref{equ4}) is stable at the point $\delta_{2^n}^{2^n}$ if and only if there exists a Lyapunov function of Boolean network (\ref{equ4}), $V_1(x(t))$, which is defined to satisfy
\begin{itemize}
\item $V_1(x(t))>0$ for $x(t)\neq\delta_{2^n}^{2^n}$ and $V_1(x(t))=0$ for $x(t)=\delta_{2^n}^{2^n}$;
\item $\Delta V_1(x(t))<0$ for $x(t)\neq\delta_{2^n}^{2^n}$ and $\Delta V_1(x(t))=0$ for $x(t)=\delta_{2^n}^{2^n}$, where $\Delta V_1(x(t)):=V_1(x(t+1))-V_1(x(t))$.
\end{itemize}
Then by (9) in \cite{wang2012definition}, a Lyapunov function for Boolean network can be constructed as
\begin{align}\label{equ5}
V_1(x(t))=\lambda^Tx(t),
\end{align} where $\lambda=(\lambda_1,0)^T\in\mathbb{R}^{2^n}$ with $\lambda_1\in\mathbb{R}^{2^n-1}$ satisfying
\begin{align}
\lambda_1&>0,\label{equ6}\\
F_{11}^T\lambda_1-\lambda_1&<0.\label{equ7}
\end{align}
\subsection{Lyapunov function for probabilistic Boolean networks}\label{sub2.3}
If the update strategy of a Boolean network is not deterministic and belongs to a set of possible update strategies with certain probability distribution, then the Boolean network becomes a probabilistic Boolean network. Consider a probabilistic Boolean network with $n$ nodes and $s$ possible update strategies as
\begin{align}\label{equ8}
Y(t+1)=g_{(t)}(Y(t)),
\end{align} where $Y(t)\in\mathbb{B}^n$, and $g_{(t)}\in\{g_1,g_2,\ldots,g_s\}$ with $g_i:\mathbb{B}^n\rightarrow\mathbb{B}^n$ being a Boolean vector function, $i=1,2,\ldots,s$. Moreover, for every time $t$, ${\bf Pr}\{g_{(t)}=g_i\}=p_i$, where $p_i\geq0$ and $\sum_{i=1}^sp_i=1$. Without loss of the generality, it is assumed that $p_i>0$ for every $i=1,2,\ldots,s$ since if $p_{i_0}=0$, then we can consider the possible update strategy set as $\{g_1,g_2,\ldots,g_s\}\backslash\{g_{i_0}\}$. The probability distribution of $g_{(t)}$ is independent of the historical states $Y(k)$ for $k\leq t$. Similar to that of the Boolean network case, by Lemma \ref{lemma1}, the equivalent algebraic form of probabilistic Boolean network (\ref{equ8}) can be obtained as
\begin{align}\label{equ9}
y(t+1)=G{(t)}y(t),
\end{align} where $y(t)=\phi_n(Y(t))\in\Delta_{2^n}$,  $G{(t)}\in\{G_1,G_2,\ldots,G_s\}$ where $G_i\in{\cal L}_{2^n\times 2^n}$ is the corresponding transition matrix of $g_{i}$, and ${\bf Pr}\{G{(t)}=G_i\}=p_i$, $i=1,2,\ldots,s$. Before constructing a Lyapunov function for (\ref{equ9}), the definitions of stochastic stability and the Lyapunov function for (\ref{equ9}) are first given as follows.
\begin{definition}\label{def1}
Probabilistic Boolean network (\ref{equ9}) is said to be stochastically stable at $\delta_{2^n}^{2^n}$ if $\lim_{t\rightarrow\infty}{\bf E}\{y(t)\}=\delta_{2^n}^{2^n}$.
\end{definition}
\begin{remark}
The definition of stochastic stability for probabilistic Boolean networks in Definition \ref{def1} is different from that in \cite{li2014state}, where a probabilistic Boolean network is said to be stable at $\delta_{2^n}^{2^n}$ with probability one if for any initial value $y(0)$, there exists an integer $T>0$ such that for all $t\geq T$, one has
\begin{align}
{\rm\bf Pr}\{y(t)=\delta_{2^n}^{2^n}|y(0)\}=1.
\end{align}
The above definition is also called finite-time stability with probability one \cite{li2019finite}. This can be regarded as special case of Definition \ref{def1}.
\end{remark}

\begin{definition}\label{def2}
A stochastic function $V_2:\Delta_{2^n}\rightarrow\mathbb{R}$ is called a Lyapunov function for probabilistic Boolean network (\ref{equ9}) if the following conditions hold:
\begin{itemize}
\item $V_2(y(t))>0$ for $y(t)\neq\delta_{2^n}^{2^n}$ and $V_2(y(t))=0$ for $y(t)=\delta_{2^n}^{2^n}$;
\item $\Delta V_2(y(t))<0$ for $y(t)\neq\delta_{2^n}^{2^n}$ and $\Delta V_2(y(t))=0$ for $y(t)=\delta_{2^n}^{2^n}$, where $\Delta V_2(y(t))={\bf E}\{V_2(y(t+1))|y(t)\}-V_2(y(t))$.
\end{itemize}
\end{definition}
\begin{lemma}\label{lemma2}
Based on Definitions \ref{def1} and \ref{def2}, probabilistic Boolean network (\ref{equ9}) is stochastically stable at $\delta_{2^n}^{2^n}$ if and only if there exists a Lyapunov function of network (\ref{equ9}).
\end{lemma}

{\em Proof.} Necessity. By Definition \ref{def1}, if probabilistic Boolean network (\ref{equ9}) is stochastically stable at $\delta_{2^n}^{2^n}$, then for any initial state $y(0)$, $\lim_{t\rightarrow\infty}{\bf E}\{y(t)\}=\delta_{2^n}^{2^n}$. Taking expectation on both side of (\ref{equ9}), one has
\begin{align}
{\bf E}\{y(t+1)\}&=\sum_{i=1}^sp_iG_i{\bf E}\{y(t)\}\nonumber\\
&:=G{\bf E}\{y(t)\},\label{equ10}
\end{align} where $G=\sum_{i=1}^sp_iG_i$ is a nonnegative matrix. Then ${\bf 1}^T_{2^n}{\bf E}\{y(t)\}=1$ and
\begin{align}\label{e11}
{\bf 1}^T_{2^n}G=\sum_{i=1}^sp_i{\bf 1}^T_{2^n}G_i=\sum_{i=1}^sp_i{\bf 1}^T_{2^n}={\bf 1}^T_{2^n}
 \end{align} as $y(t)\in\Delta_{2^n}$ and $G_i\in{\cal L}_{2^n\times 2^n}$, $i=1,2,\ldots,s$. Partition $G$ and  ${\bf E}\{y(t)\}$, respectively, as
\begin{align*}
G=\left[\begin{array}{cc}G_{11}&G_{12}\\G_{21}&G_{22}\end{array}\right],~~{\bf E}\{y(t)\}=\left[\begin{array}{c}\tilde{y}_1(t)\\ \tilde{y}_2(t)\end{array}\right],
\end{align*} where $G_{11}\in\mathbb{R}^{(2^n-1)\times(2^n-1)}$, $G_{22}\in\mathbb{R}$, $\tilde{y}_1(t)\in\mathbb{R}^{2^n-1}$ and $\tilde{y}_2(t)\in\mathbb{R}$. Thus, $\lim_{t\rightarrow\infty}{\bf E}\{y(t)\}=\delta_{2^n}^{2^n}$ if and only if $\lim_{t\rightarrow\infty}\tilde{y}_1(t)={\bf 0}_{2^n-1}$ and $\lim_{t\rightarrow\infty}\tilde{y}_2(t)=1$.
Taking limitation on both side of (\ref{equ10}) yields $\lim_{t\rightarrow\infty}{\bf E}\{y(t+1)\}=G\lim_{t\rightarrow\infty}{\bf E}\{y(t)\}$, i.e., $\delta_{2^n}^{2^n}=G\delta_{2^n}^{2^n}$, which implies $G_{12}={\bf0}_{2^n-1}$ and $G_{22}=1$. Then from (\ref{equ10}), the update of $\tilde{y}_1(t)$ can be written as
\begin{align}\label{equ11}
\tilde{y}_1(t+1)=G_{11}\tilde{y}_1(t).
\end{align} Thus, $\lim_{t\rightarrow\infty}\tilde{y}_1(t)={\bf 0}_{2^n-1}$ if and only if $\rho(G_{11})<1$. Note that the matrix $G_{11}$ is nonnegative. By \cite{horn2012matrix}, $\lim_{t\rightarrow\infty}\tilde{y}_1(t)={\bf 0}_{2^n-1}$ if and only if there exists a vector $\nu_1\in\mathbb{R}^{2^n-1}$ such that
\begin{align}
\nu_1&>0,\label{equ12}\\
G_{11}^T\nu_1-\nu_1&<0.\label{equ13}
\end{align}
Define
\begin{align}\label{equequ}
V_2(y(t))=\nu^Ty(t),
\end{align} where $\nu=(\nu_1^T,0)^T\in\mathbb{R}^{2^n}$ with $\nu_1$ satisfying (\ref{equ12}) and (\ref{equ13}).
It can be easily verified that $V_2(y(t))$ matches the conditions in Definition \ref{def2} and thus can be viewed as a Lyapunov function of network (\ref{equ9}).

Sufficiency. If there exists a Lyapunov function $V_2(y(t))$ in the form (\ref{equequ}) satisfying the conditions in Definition \ref{def2}, then (\ref{equ12}), (\ref{equ13}) hold and $G_{12}={\bf 0}_{2^n-1}$. With the necessity proof, one can prove that the network (\ref{equ9}) is stochastically stable at $\delta_{2^n}^{2^n}$. \hfill$\blacksquare$

From the proof of Lemma \ref{lemma2}, a Lyapunov function of network (\ref{equ9}) can be constructed as
\begin{align}
V_2(y(t))=\nu^Ty(t),
\end{align} where $\nu=(\nu_1^T,0)^T\in\mathbb{R}^{2^n}$ with $\nu_1\in\mathbb{R}^{2^n-1}$ satisfying
\begin{align}
\nu_1&>0,\label{}\\
G_{11}^T\nu_1-\nu_1&<0.\label{}
\end{align}

\subsection{Lyapunov function for Markovian switching Boolean networks}\label{sub2.4}
If the update strategy at every time randomly chooses the possible update strategies related to the one at the last time rather than following a certain probabilistic distribution, then this network can be modeled as a Markovian switching Boolean network given as
\begin{align}\label{equ15}
Z(t+1)=h_{\sigma(t)}(Z(t)),
\end{align} where $Z(t)\in\mathbb{B}^n$, $h_{\sigma(t)}\in\{h_1,h_2,\ldots,h_r\}$ with $h_i:\mathbb{B}^n\rightarrow\mathbb{B}^n$ being a Boolean vector function, $\sigma(t)$ is a switching signal, which is a discrete-time homogeneous Markov chain with finite state set ${\cal R}=\{1,2,\ldots,r\}$, i.e., $\sigma(t)\in{\cal R}$, and its transition probability matrix as $\Pi=(\pi_{ij})\in\mathbb{R}^{r\times r}$ defined as
\begin{align*}
\pi_{ij}={\bf Pr}\{\sigma(t+1)=j|\sigma(t)=i\},
\end{align*} where $\pi_{ij}\geq 0$ for $i,j\in{\cal R}$ and $\sum_{j=1}^r\pi_{ij}=1$ for any $i\in{\cal R}$. The algebraic form of network (\ref{equ15}) is
\begin{align}\label{equ16}
z(t+1)=H_{\sigma(t)}z(t),
\end{align} where $z(t)=\phi_n(Z(t))\in\Delta_{2^n}$ and $H_{\sigma(t)}\in\{H_1,H_2,\ldots,H_{r}\}$ where $H_i\in{\cal L}_{2^n\times 2^n}$ is the transition matrix corresponding to $h_i$, $i=1,2,\ldots,r$.

As in \cite{meng2017stability,meng2018stability}, it is assumed that the Markov chain $\sigma(t)$ is ergodic, i.e., irreducible and positive recurrent.
\begin{definition}[\cite{meng2017stability,meng2018stability}]
Markovian switching Boolean network (\ref{equ16}) is said to be stochastically stable at $\delta_{2^n}^{2^n}$ if for any initial value $z(0)$ and
any initial distribution of $\sigma(t)$, the following condition holds:
\begin{align}
\lim_{t\rightarrow\infty}{\rm\bf E}\{z(t)|z(0),\sigma(0)\}=\delta_{2^n}^{2^n}.
\end{align}
\end{definition}

For network (\ref{equ16}), the Lyapunov function is defined as follows.
\begin{definition}[\cite{meng2017stability}]\label{def5}
A stochastic function $V_3:\Delta_{2^n}\times{\cal R}\rightarrow\mathbb{R}$ is called a Lyapunov function of network (\ref{equ16}) if for any $\sigma(t)\in{\cal R}$,
\begin{itemize}
\item $V_3(z(t),\sigma(t))>0$ for $z(t)\neq \delta_{2^n}^{2^n}$, and $V_3(z(t),\sigma(t))=0$ for $z(t)=\delta_{2^n}^{2^n}$;
\item $\Delta V_3(z(t),\sigma(t))<0$ for $z(t)\neq \delta_{2^n}^{2^n}$, and $\Delta V_3(z(t),\sigma(t))=0$ for $z(t)=\delta_{2^n}^{2^n}$, where $\Delta V_3(z(t),\sigma(t))={\bf E}\{V_3(z(t+1),\sigma(t+1))| z(t),\sigma(t)\}-V_3(z(t),\sigma(t))$.
\end{itemize}
\end{definition}

It has also been proved in \cite{meng2017stability} that Markovian switching Boolean network (\ref{equ16}) is stochastically stable at $\delta_{2^n}^{2^n}$ if and only if there exists a Lyapunov function for (\ref{equ16}) defined in Definition \ref{def5}. Note that $H_{\sigma(t)}=H_i$ when $\sigma(t)=i$. Partition $H_i$ as
\begin{align}
H_i=\left[\begin{array}{cc}H_{i,11}&H_{i,12}\\H_{i,21}& H_{i,22}\end{array}\right], ~~H_{i,11}\in\mathbb{B}^{(2^n-1)\times(2^n-1)},
\end{align} for $i=1,2,\ldots,r$. By recalling the stability results in \cite{meng2017stability,meng2018stability},  a Lyapunov function for Markovian switching Boolean network (\ref{equ16}) can be designed as
\begin{align}
V_3(z(t),\sigma(t))=\omega_{\sigma(t)}^Tz(t),
\end{align} where $\sigma(t)\in{\cal R}$ and $\omega_i=(\omega_{i1}^T,0)^T\in\mathbb{R}^{2^n}$ with $\omega_{i1}\in\mathbb{R}^{2^n-1}$ satisfying
\begin{align}
\sum_{j=1}^r\pi_{ij}H_{i,11}^T\omega_{j1}-\omega_{i1}&<0,\\
\omega_{i1}&>0,
\end{align} for $i=1,2,\ldots,r$.
\begin{remark}
For deterministic Boolean networks, a deterministic function, of course, can be regarded as a Lyapunov function. The Lyapunov functions for probabilistic and Markovian switching Boolean networks are both stochastic, while a Lyapunov function for a probabilistic Boolean network can be equipped with a common gain $\nu$ and a Lyapunov function for a Markovian switching Boolean network is in fact composed of multiple functions. It is also difficult to find a common Lyapunov function for a Markovian switching Boolean network.
\end{remark}

\section{Self-triggered scheduling}\label{section3}
To reduce computational loads and deal with the cases with constrained resources,  we aim to design self-triggered strategy to properly set up the rules for updating the system control when necessary. In fact, the control strategy under self-triggered case has the following structure:
\begin{equation}\label{equ22}
\left\{\begin{array}{ll}
u(t)=u(t_k)\in{\cal U}(x(t_k)),&t\in[t_k,t_{k+1}),\\
t_{k+1}=t_k+\tau(x(t_k)),
\end{array}
\right.
\end{equation} where $t_0=0$, $\tau(x(t_k))$ denotes the time between two consecutive sampling times, and ${\cal U}(x(t_k))$ is the possible control set when the state is $x(t_k)$. The problem we are interested is to solve the co-design problem of both the triggering times and the required control.

In this section, self-triggered control for BCNs, probabilistic BCNs and Markovian switching BCNs will be studied mainly based on the Lyapunov theory presented in the previous section. Hereafter, it is assumed that a feedback controller is given such that the studied BCN can be (stochastically) stabilizable at the equilibrium point $\delta_{2^n}^{2^n}$ since the stabilization of BCNs can be viewed as a prior by using the existing methods in \cite{Cheng2011,li2013state,wang2019stabilization,meng2018stability}. The notations in Subsections \ref{sub4.1}, \ref{sub4.2} and \ref{sub4.3} are also independent.

\subsection{BCNs}\label{sub4.1}
In this subsection, we just study the BCN from its algebraic expression form as
\begin{align}\label{equ23}
x(t+1)=Fu(t)x(t),
\end{align} where $x(t)\in\Delta_{2^n}$ is the state variable, $u(t)\in\Delta_{2^m}$ is the control input and $F\in{\cal L}_{2^n\times 2^{n+m}}$. Assume that BCN (\ref{equ23}) is stabilizable at $\delta_{2^n}^{2^n}$ by a state feedback control
 \begin{align}\label{equ24}
 u(t)=Kx(t),
 \end{align} where $K\in{\cal L}_{2^m\times 2^n}$, then by recalling the Lyapunov function in Subsection \ref{sub2.2}, there exists a Lyapunov function $V_1(x(t))=\lambda^Tx(t)$ for the closed-loop system
 \begin{align}\label{equ25}
 x(t+1)=FK\Phi_{2^n}x(t),
 \end{align}  satisfying
\begin{align*}
\lambda=(\lambda_1^T,0)^T,~~\lambda_1>0,~~\tilde{F}_{11}^T\lambda_1-\lambda_1<0, ~~\lambda_1\in\mathbb{R}^{2^n-1},
\end{align*} where $$\tilde{F}_{11}=[I_{2^n-1} ~{\bf 0}_{2^n-1}]FK\Phi_{2^n}\left[\begin{array}{l}I_{2^n-1}\\{\bf0}_{1\times(2^n-1)}\end{array}\right],$$ and $\Phi_{2^n}=\text{diag}(\delta_{2^n}^1,\delta_{2^n}^2,\ldots,\delta_{2^n}^{2^n})$ is called a reduced order matrix \cite{Chengd2011} such that $\Phi_{2^n}x(t)=x(t)\ltimes x(t)$.
 By recalling the Lyapunov function in Subsection \ref{sub2.2}, the self-triggered scheduling is designed such that the Lyapunov function at the next time will decrease.
For $M\geq1$ and a state $x\in\Delta_{2^n}$, if for any $t$ and any $u\in\Delta_{2^m}$, $(Fu)^tx\neq\delta_{2^n}^{2^n}$, let
\begin{align}
&{\cal U}_M(x):=\{u\in\Delta_{2^m}\mid V_1\left((Fu)^{i}x\right)-V_1((Fu)^{i-1}x)<0,\nonumber\\&~~~~~~~~~~~~~~~i=1,2,\ldots,M \}.\label{equequequ26}
\end{align} Otherwise, if there exist some $u\in\Delta_{2^m}$ and a positive integer $N_u\leq M$ such that $(Fu)^{N_u}x=\delta_{2^n}^{2^n}$, denote
\begin{align}
&{\cal U}_M(x):=\{u\in\Delta_{2^m}\mid V_1\left((Fu)^{i}x\right)-V_1((Fu)^{i-1}x)<0,\nonumber\\
&~~~~~~~~~~~~~V_1\left((Fu)^{j}x\right)-V_1((Fu)^{j-1}x=0,\nonumber\\
&~~~~~~~~~~~~~i=1,2,\ldots,N_u; j=N_u+1,\ldots,M \}.
\end{align}
 Then
 $\tau(x(t_k))$ and ${\cal U}(x(t_k))$ in (\ref{equ22}) are defined formally as
\begin{align}
&\tau(x(t_k))=\max\{M\mid {\cal U}_M(x(t_k))\neq\emptyset\},\label{equequ28}\\
&{\cal U}(x(t_k))={\cal U}_{\tau(x(t_k))}(x(t_k)).\label{equequ29}
\end{align}
\begin{theorem}\label{thm1}
Consider BCN (\ref{equ23}). The control strategy (\ref{equequ28}), (\ref{equequ29}) for BCN (\ref{equ23}) is well defined, i.e., for all $x(0)\in\Delta_{2^n}$, $t_{k+1}>t_k$, for $k=1,2,\ldots$, and there exists a positive integer $N<2^n$ such that for any $t\geq t_N$, the control $u(t)$ remains unchanged, i.e., $t_N<\infty$ and $t_{N+1}=\infty$. Moreover, the system (\ref{equ23}) with the control strategy (\ref{equ22}) is stabilizable at $\delta_{2^n}^{2^n}$ in a finite time.
\end{theorem}

{\em Proof}. To show the well-definedness of the control strategy (\ref{equequ28}), (\ref{equequ29}), it suffices to prove that for all $x\in\Delta_{2^n}$, ${\cal U}_1(x)\neq\emptyset$ where ${\cal U}_1(x)$ is defined in (\ref{equequequ26}).
  Suppose that at some sampling time $t_k$, $x(t_k)=x$. Choosing $\bar{u}=Kx$, where $K$ is given in (\ref{equ24}), we have
\begin{align*}
V_1(F\bar{u}x)-V_1(x)&=V_1(FK\Phi_{2^n}x(t_k))-V_1(x(t_k))\\
&=V_1(x(t_k+1))-V_1(x(t_k))\\
&\left\{\begin{array}{ll}=0,&{\rm if}~x=\delta_{2^n}^{2^n},\\<0,&{\rm if}~x\neq\delta_{2^n}^{2^n},\end{array}\right.
\end{align*} by the definition of Lyapunov function $V_1(x(t))$. Then $\bar{u}\in{\cal U}_1(x)$. This proves that ${\cal U}_1(x)\neq \emptyset$, and thus $t_{k+1}>t_k$.

Now we are in a position to prove that there exists a positive integer $N<2^n$ such that  the  update of the control $u(t)$ stops at $t_N$, i.e. $t_N<\infty$ and $t_{N+1}=\infty$. Bearing in mind the self-triggered scheduling in (\ref{equequ28}) and (\ref{equequ29}), we have $V_1(x(t))>V_1(x(t+1))$ if $x(t)\neq \delta_{2^n}^{2^n}$ and $V_1(x(t))=V_1(x(t+1))=\cdots=0$ otherwise.
Note that the number of all the possible values of the Lyapunov function $V_1(x(t))$ for a fixed a $\lambda$ is no more than $2^n$ since $x(t)\in\Delta_{2^n}$.
If $N\geq2^n$, then by the definition of the Lyapunov function $V_1(x(t))$,  one can find an integer $i$ satisfying $0\leq i< N$ such that $V_1(x(t_i))=0$, i.e. $x(t_i)=\delta_{2^n}^{2^n}$. By selecting $u(t)=Kx(t_i)$ for all $t\geq t_i$, where $K$ is given in (\ref{equ24}), then $(Fu(t_i))^tx(t_i)=\delta_{2^n}^{2^n}$ for any $t\geq 0$. That is, for any $t$, $V_1(x(t))=0$ for all $t\geq t_i$. Then the control will not update after $t_i$, i.e., $t_{i+1}=\infty$, which is a contradiction to $t_N<\infty$ and $N>i$.

Next, we prove that the  system (\ref{equ23}) with the control strategy (\ref{equequ28}), (\ref{equequ29}) reaches the stable point $\delta_{2^n}^{2^n}$ at a finite time and remains unchanged. By the self-triggered condition, one has that $V_1(t_0)>V_1(t_1)>\cdots>V_1(t_N)\geq V_1(t_{N+1})=0$. If $x(t_N)=\delta_{2^n}^{2^n}$, then $u(t)=Kx(t_N)$ for all $t\geq t_N$ can guarantee that the state of the system (\ref{equ23}) is $\delta_{2^n}^{2^n}$ afterwards. That is to say the system (\ref{equ23}) is stabilizable at $\delta_{2^n}^{2^n}$ in finite time $t_N$. If $x(t_N)\neq\delta_{2^n}^{2^n}$, then $u(t)=Kx(t_N)$ for all $t\geq t_N$ can guarantee that the state of the system (\ref{equ23}) reaches $\delta_{2^n}^{2^n}$ in time $2^n$ since the state space of a Boolean network is finite \cite{Cheng2011,li2013state}.  Therefore the system (\ref{equ23}) is stabilizable at $\delta_{2^n}^{2^n}$ in finite time $t_N+2^n$. The proof is completed.
\hfill$\blacksquare$
\begin{remark}
From the above analysis, self-triggered controllers are not unique and can also be designed based on the decrease of the Lyapunov function.  After $t_{k+1}$, at which the control should be updated, is determined, the state at time $t_{k+1}$ and the possible control set ${\cal U}(x(t_{k+1}))$ can also be computed. The control at time $t_{k+1}$ can be chosen from the possible control set such that the Lyapunov function $V(x(t))$ take the smallest value at $t_{k+1}+1$. The detailed control design process is given as follows.
Define $${\cal I}(x(t_{k+1}))=\arg\min\limits_{u\in{\cal U}(x(t_{k+1}))}\left\{\lambda^TFux(t_{k+1})\right\}.$$ Then the corresponding self-triggered  controller can be given as
\begin{align}
\left\{\begin{array}{l}u(t)=u(t_k), ~\text{for}~t\in[t_k,t_{k+1});\\
u(t)\in{\cal I}(x(t_{k+1})),~\text{for}~t=t_{k+1}.\end{array}\right.
\end{align}
\end{remark}
\begin{example}
Consider a BCN with $n=3$, $m=1$ and the transition matrix in (\ref{equ23}) as $$F=\delta_8[2,3,3,3,7,7,8,8,4,4,6,6,8,8,5,5].$$ A feasible state feedback controller is pre-given as $K=\delta_2[1,1,2,2,2,1,2,1]$. Then a Lyapunov function exists in the form as $V(x(t))=\lambda^Tx(t)$ with $\lambda=(\lambda_1,\lambda_2,\lambda_3,\lambda_4,\lambda_5,\lambda_6,\lambda_7,0)^T$ being selected as $\lambda_1=5$, $\lambda_2=4.5$, $\lambda_3=4$, $\lambda_4=6$, $\lambda_5=1$, $\lambda_6=3$, $\lambda_7=2$.
Consequently, the self-triggered scheduling can be designed based on the Lyapunov stability theory. Specifically, at $t_0=0$, if $x(t_0)=\delta_8^1$, no matter $u=\delta_2^1$ or $u=\delta_2^2$, $(Fu)^tx(t_0)$ will not be $\delta_{8}^8$ for any $t$. By a simple computation, $\tau(x(t_0))$ and ${\cal U}(x(t_0))$ in (\ref{equequ28}) and (\ref{equequ29}) are $\tau(x(t_0))=2$ and ${\cal U}(x(t_0))=\{\delta_2^1\}$. Therefore, $u(t_0)=\delta_2^1$, and the next triggering time is $t_1=t_0+\tau(x(t_0))=2$. Note that at time $t_1=2$, the state $x(t_1)$ is $\delta_8^3$. Similarly, one can compute that $\tau(x(t_1))=2$ and ${\cal U}(x(t_1))=\{\delta_2^2\}$. Thus, $u(t_1)=\delta_2^2$ and the next triggering time is $t_2=t_1+\tau(x(t_1))=4$. Here, $x(t_2)$ is the equilibrium point $\delta_8^8$. Let $u(t_2)=K\delta_8^8=\delta_2^1$, then the next triggering time is $t_3=\infty$.
 Therefore, it only needs the sampled data and control strategies at three times $t=0$, $t=2$ and $t=4$ to ensure the studied BCN stable at $\delta_8^8$.
 Note that the self-triggered sampling times are related with the initial states. Fortunately, all the sampling times according the mechanism corresponding to different initial states can be designed when performing the self-triggered scheduling (see Table 1).
 \begin{table}[ht]
 \centering
 \caption{Sampling scheduling}
\begin{tabular}{| c|c |c |}
  \hline
  initial states &  sampling times &control\\
    \hline
  $\delta_8^1$ & $t_0=0,t_1=2$, & $u(t_0)=\delta_2^1,u(t_1)=\delta_2^2,$\\
  &$t_2=4$&$u(t_2)=\delta_2^1$\\
  $\delta_8^2$ & $t_0=0,t_1=1$& $u(t_0)=\delta_2^1,u(t_1)=\delta_2^2$ \\
  &$t_2=3$&$u(t_2)=\delta_2^1$\\
  $\delta_8^3$ & $t_0=0,t_1=2$& $u(t_0)=\delta_2^2,u(t_1)=\delta_2^1$\\
  $\delta_8^4$ & $t_0=0,t_1=2$& $u(t_0)=\delta_2^2,u(t_1)=\delta_2^1$\\
   $\delta_8^5$ & $t_0=0,t_1=1$& $u(t_0)=\delta_2^2,u(t_1)=\delta_2^1$\\
    $\delta_8^6$ & $t_0=0$& $u(t_0)=\delta_2^1$\\
     $\delta_8^7$ & $t_0=0$& $u(t_0)=\delta_2^1$\\
     $\delta_8^8$ & $t_0=0$& $u(t_0)=\delta_2^1$\\
  \hline
\end{tabular}
\end{table}
\end{example}
\subsection{Probabilistic BCNs}\label{sub4.2}
Consider a probabilistic BCN as
\begin{align}\label{equ29}
y(t+1)=G{(t)}u(t)y(t),
\end{align} where $y(t)\in\Delta_{2^n}$ is the state variable, $u(t)\in\Delta_{2^m}$ is control input, and $G{(t)}\in\{G_1,G_2,\ldots,G_s\}$ with $G_i\in{\cal L}_{2^n\times2^{n+m}}$. Moreover, ${\bf Pr}\{G{(t)}=G_i\}=p_i$, where $p_i>0$ and $\sum_{i=1}^sp_i=1$. Assume that  probabilistic BCN (\ref{equ29}) is stabilizable by state feedback controller
\begin{align}\label{equ30}
u(t)=Ky(t),
\end{align} where $K\in{\cal L}_{2^m\times 2^n}$. Then the closed-loop system
\begin{align}\label{equ31}
y(t+1)=G{(t)}K\Phi_{2^n}y(t)
\end{align} is stochastically stable to $\delta_{2^n}^{2^n}$. By taking expectation on both sides of (\ref{equ31}), one has
\begin{align}\label{equ32}
{\bf E}\{y(t+1)\}=\sum_{i=1}^sp_iG_{i}K\Phi_{2^n}{\bf E}\{y(t)\}:=\tilde{G}{\bf E}\{y(t)\},
\end{align} where $\tilde{G}:=\sum_{i=1}^sp_iG_{i}K\Phi_{2^n}$.
 Based on the Lyapunov function for probabilistic Boolean networks in Subsection \ref{sub2.3}, there exists a Lyapunov function $V_2(y(t))=\nu^Ty(t)$ for the closed-loop system (\ref{equ32}) satisfying
\begin{align*}
\nu=(\nu_1^T,0)^T,~~\nu_1>0,~~\tilde{G}_{11}^T\nu_1-\nu_1<0,~~\nu_1\in\mathbb{R}^{2^n-1},
\end{align*} where $$\tilde{G}_{11}=[I_{2^n-1}~{\bf0}_{2^n-1}]\tilde{G}\left[\begin{array}{c}I_{2^n-1}\\{\bf0}_{1\times(2^n-1)}\end{array}\right].$$ Then the self-triggered scheduling (\ref{equ22}) for probabilistic BCN (\ref{equ29}) can be desiged as follows. For $M>0$, if for $u\in\Delta_{2^m}$ and any $t$, ${\bf E}\{y_{u,t}(t_k)|y(t_k)\}\neq\delta_{2^n}^{2^n}$, denote
\begin{align}
&{\cal U}_M(y(t_k))
=\{u\in\Delta_{2^m}\mid{\bf E}\{V_2\left(y_{u,i}(t_k)\right)|y(t_k)\}\nonumber\\&~~~~-{\bf E}\{V_2(y_{u,i-1}(t_k))|y(t_k)\}<0, i=1,2,\ldots,M\},\label{equ35}
\end{align} where $y_{u,M}(t_k)=(G{(t_k+M-1)}u)\cdots(G{(t_k)}u)y(t_k)$ and $y_{u,0}(t_k)=y(t_k)$. Otherwise, if there exist some $u\in\Delta_{2^m}$ and a positive integer $N_u\leq M$ such that ${\bf E}\{y_{u,N_u}(t_k)|y(t_k)\}=\delta_{2^n}^{2^n}$, denote
\begin{align}
&{\cal U}_M(y(t_k))
=\{u\in\Delta_{2^m}\mid\nonumber\\
&~~~~{\bf E}\{V_2\left(y_{u,i}(t_k)\right)|y(t_k)\}-{\bf E}\{V_2(y_{u,i-1}(t_k))|y(t_k)\}<0,\nonumber\\
 &~~~~{\bf E}\{V_2\left(y_{u,j}(t_k)\right)|y(t_k)\}-{\bf E}\{V_2(y_{u,j-1}(t_k))|y(t_k)\}=0,\nonumber\\
 &~~~~ i=1,2,\ldots,N_u, j=N_u+1,\ldots,M\}.\label{equequequ36}
\end{align}

 Also, $\tau(y(t_k))$ and ${\cal U}(y(t_k))$ are defined as
\begin{align}
&\tau(y(t_k))=\max\{M\mid{\cal U}_M(y(t_k))\neq\emptyset\},\label{equ42}\\
&{\cal U}(y(t_k))={\cal U}_{\tau(y(t_k))}(y(t_k)).\label{equ43}
\end{align}
\begin{theorem}\label{thm2}
Consider probabilistic BCN (\ref{equ29}). The control strategy in (\ref{equ42}), (\ref{equ43}) for (\ref{equ29}) is well defined, i.e., $t_{k+1}>t_k$ for $k=1,2,\ldots.$ Moreover, the system (\ref{equ29}) with the control strategy in (\ref{equ22}) is stochastically stabilizable at $\delta_{2^n}^{2^n}$.
\end{theorem}

{\em Proof.} Similar to the proof of Theorem \ref{thm1}, it suffices to prove that for all $y\in\Delta_{2^n}$, ${\cal U}_1(y)\neq\emptyset$ where ${\cal U}_1(y)$ is defined in (\ref{equ35}). It can be assumed that at some time $t_k$, $y(t_k)=y$. Let $\bar{u}=Ky$, where $K$ is the stabilizing controller given in (\ref{equ30}), then
\begin{align*}
&{\bf E}\{V_2(y_{\bar{u},1})|y\}-V_2(y)\\
&={\bf E}\{V_2(G_{t_k}K\Phi_{2^n}y(t_k))|y(t_k)\}-V_2(y(t_k))\\
&={\bf E}\{V_2(y(t_k+1))|y(t_k)\}-V_2(y(t_k))\\
&\left\{\begin{array}{ll}=0,&{\rm if}~y=\delta_{2^n}^{2^n},\\<0,&{\rm if}~y\neq\delta_{2^n}^{2^n},\end{array}\right.
\end{align*} where the last inequality is implied by the Lyapunov stability theory for probabilistic Boolean networks. Then $\bar{u}\in{\cal U}_1(x(t_k))$. Thus, $\tau(t_k)\geq1$ and the control strategy is well defined.

Now we will prove that the system (\ref{equ29}) with the control strategy in (\ref{equ22}) is stochastically stabilizable at $\delta_{2^n}^{2^n}$.
In what follows, two cases are discussed.

Case 1): There is a minimal finite time $N$ such that for any $y(0)\in\Delta_{2^n}$, ${\bf E}\{y(N)|y(0)\}=\delta_{2^n}^{2^n}$. Suppose that a maximal $k$ can be found such that $t_k< N$.
Under the control $u(t)=u(t_k)$ for any $t_k\leq t< N$, we have ${\bf E}\{y(N)|y(t_k)\}=\delta_{2^n}^{2^n}$.
Based on the sampling scheduling, for all $t\geq N$, ${\bf E}\{V_2\left(y_{u,t-N}(N)\right)|y(N)\}=0$, which is equivalent to ${\bf E}\{y_{u,t-N}(N)|y(N)\}=\delta_{2^n}^{2^n}$.
At this point, the system (\ref{equ29}) is stochastically stabilizable at $\delta_{2^n}^{2^n}$ in finite time.

Case 2):  A finite time $N$ satisfying that for any $y(0)\in\Delta_{2^n}$, ${\bf E}\{y(N)|y(0)\}=\delta_{2^n}^{2^n}$ cannot be found. Then for any time $t$ and any $y(0)\in\Delta_{2^n}$, ${\bf E}\{y(t)|y(0)\}\neq\delta_{2^n}^{2^n}$ and $V_2(y(t_k))>0$ by the definition of Lyapunov function of probabilistic Boolean networks. Therefore, for any $k$, ${\bf E}\{V_2(y(t_{k+1})|y(t_k)\}-V_2(y(t_k))<0$, based on which a sufficiently small positive number $\alpha<1$ can be found such that
\begin{align}\label{equ36}
{\bf E}\{V_2(y(t_{k+1})|y(t_k)\}<(1-\alpha) V_2(y(t_k))
\end{align} for any $k=0,1,\ldots.$ Taking expectation on both sides of (\ref{equ36}) yields
\begin{align*}
{\bf E}\{{\bf E}\{V_2(y(t_{k+1})|y(t_k)\}|y(t_0)\}\leq(1-\alpha){\bf E}\{V_2(y(t_k))|y(t_0)\},
\end{align*} that is,
\begin{align*}
{\bf E}\{V_2(y(t_{k+1})|y(t_0)\}\leq(1-\alpha){\bf E}\{V_2(y(t_k))|y(t_0)\}.
\end{align*} By iteration,
$${\bf E}\{V_2(y(t_{k})|y(t_0)\}\leq(1-\alpha)^k{\bf E}\{V_2(y(t_0))|y(t_0)\}.$$ Making $k\rightarrow\infty$ produces $\lim_{k\rightarrow\infty}{\bf E}\{V_2(y(t_{k})|y(t_0)\}=0$, which is equivalent to $\lim_{k\rightarrow\infty}{\bf E}\{y(t_k)\}=\delta_{2^n}^{2^n}$.
\hfill$\blacksquare$
%

Next we give an example on a probabilistic Boolean control network to show that its stochastic stability can be ensured by the self-triggered control strategy.
\begin{example}
Consider a probabilistic BCN in the form of (\ref{equ29}) with $n=3$, $m=1$, and ${\bf P}\{G(t)=G_1\}=p_1=0.3$, ${\bf P}\{G(t)=G_2\}=p_2=0.7$, where
\begin{align*}
&G_1=\delta_8[3,1,6,6,2,2,8,8,1,1,1,8,4,3,5,8],\\
&G_2=\delta_8[1,1,2,6,8,7,7,7,6,1,1,1,5,5,5,8].
\end{align*} A feasible update-based feedback control is given as $u(t)=Ky(t)$, where
\begin{align*}
K=\delta_2[2,2,1,1,1,1,2,2].
\end{align*} Then a feasible Lyapunov function can be given as $V_2(y(t))=\nu^Ty(t)$, where $\nu=(8.3, 9.3, 9.4,6.5,2.8,6.4,3.6,0)^T$. Via the obtained results in this subsection, the self-triggered scheduling (\ref{equ42}), (\ref{equ43}) can be performed by MATLAB with the simulation results being shown in Figure 1. In Figure 1(a), we take the initial state $y(0)$ as $y(0)=\delta_8^1$ and the corresponding state trajectories are given by running the program 500 times. In Figure 1(b), the possible trajectories corresponding to all initial states are simulated. From these, it can also be seen that the stochastic stability at $\delta_8^8$ can be ensured.

\begin{figure}[htbp]
\centering
\subfigure[]{
\centering
\includegraphics[width=2.5in]{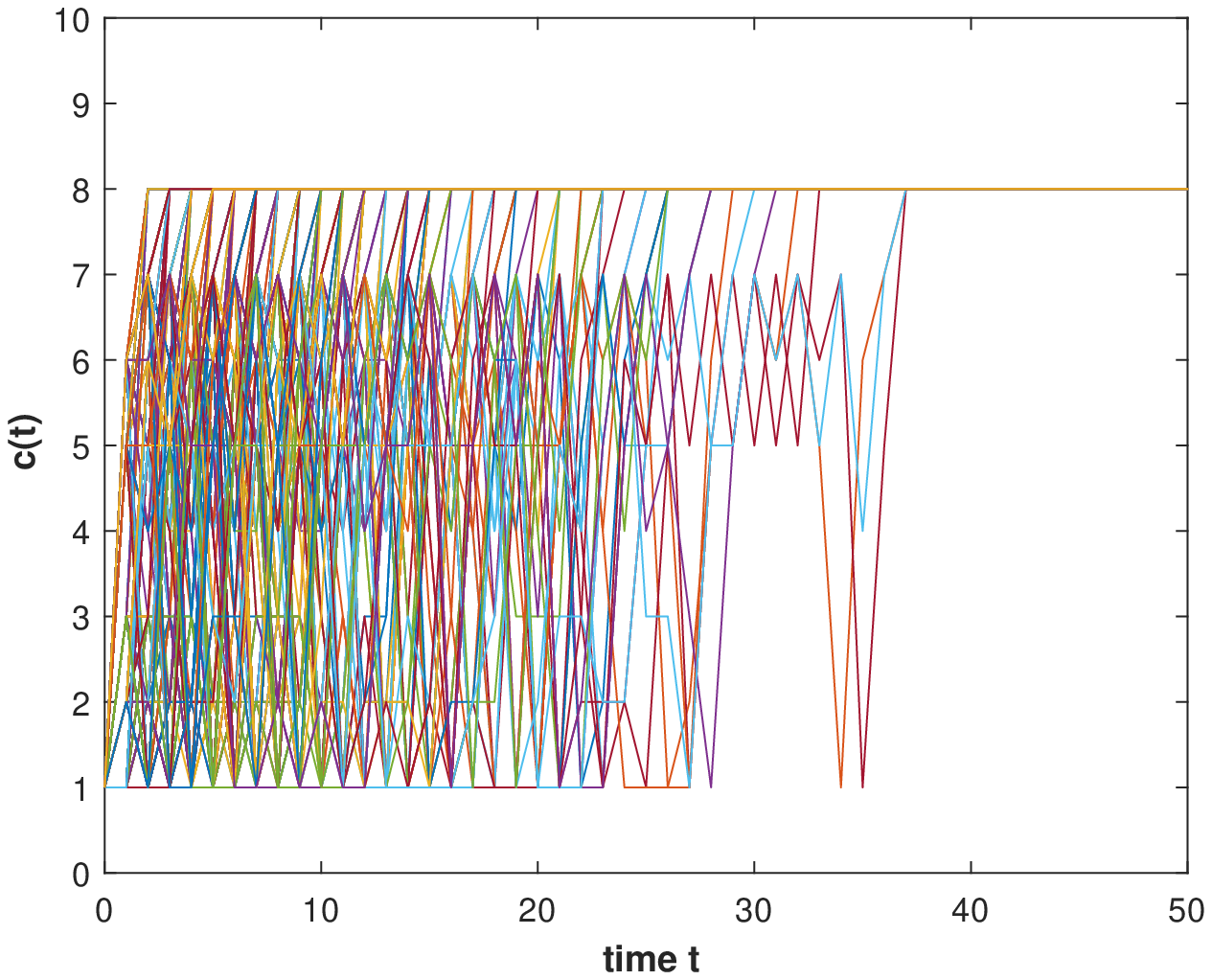}
}
\subfigure[]{
\centering
\includegraphics[width=2.5in]{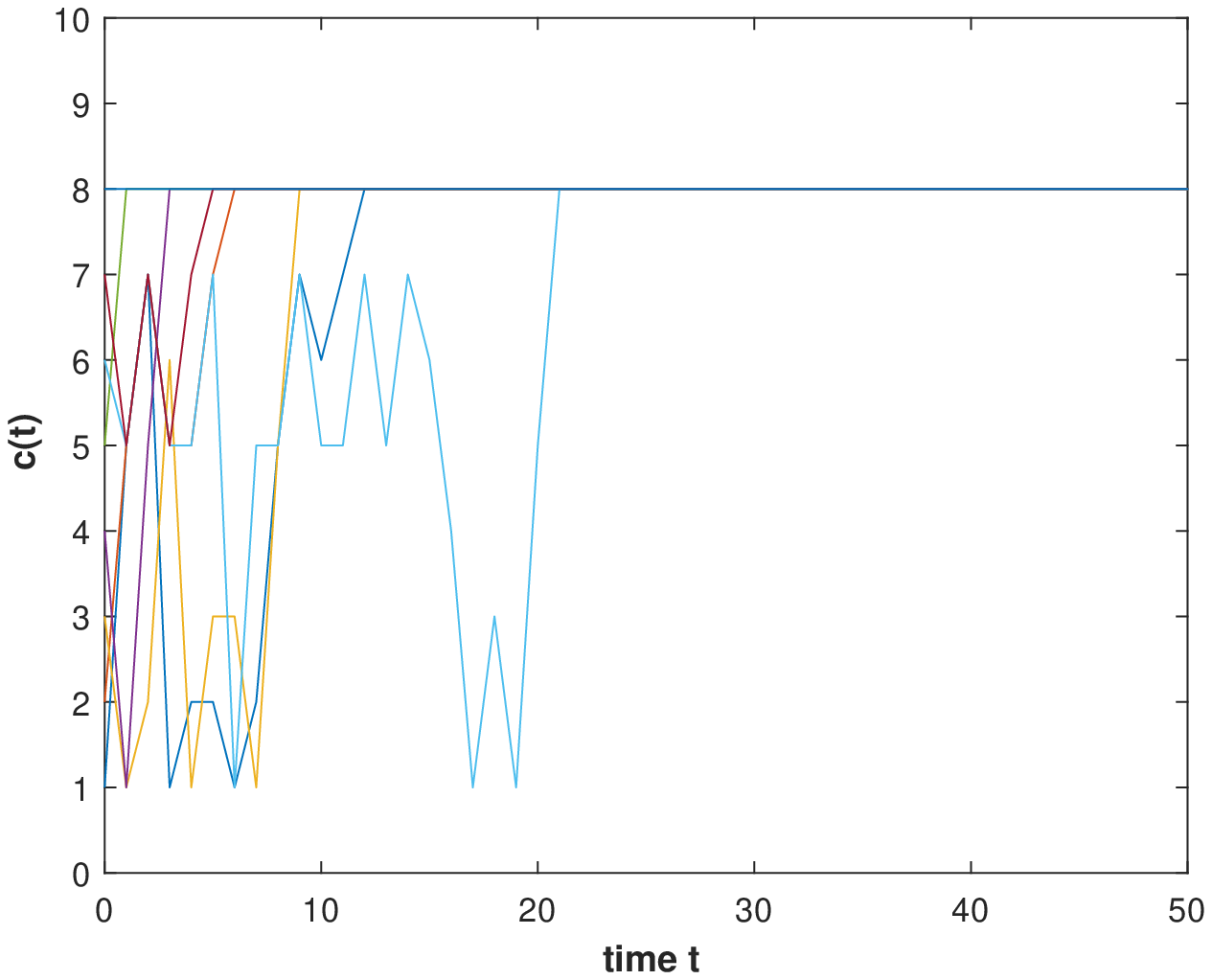}
}%
\centering
\caption{(a) The trajectories corresponding to initial state $y(0)=\delta_8^1$ by running the program 500 times and (b) The possible trajectories corresponding to all the initial states. $c(t)$ is the index of 1 in $y(t)$, i.e., $y(t)=\delta_8^{c(t)}$.}
\end{figure}

%

\end{example}
\subsection{Markovian switching BCNs}\label{sub4.3}
Consider a Markovian switching BCN as
\begin{align}\label{equ37}
z(t+1)=H_{\sigma(t)}u(t)z(t),
\end{align} where $z(t)\in\Delta_{2^n}$ is the state variable, $u(t)\in\Delta_{2^m}$ is the control input, $\sigma(t)$ is the  switching signal, and $H_{\sigma(t)}\in\{H_1,H_2,\ldots,H_r\}$ with $H_i\in{\cal L}_{2^n\times 2^{n+m}}$, $i=1,2,\ldots,r$. Here $\sigma(t)$ is a discrete Markov chain same as in Subsection \ref{sub2.4}. If Markovain switching BCN (\ref{equ37}) is stochastically stabilizable at $\delta_{2^n}^{2^n}$ by a state feedback control
\begin{align}\label{equ38}
u(t)=Kx(t),
\end{align} where $K\in{\cal L}_{2^m\times 2^n}$, then the closed-loop system
\begin{align}\label{equ39}
z(t+1)=H_{\sigma(t)}K\Phi_{2^n}z(t)
\end{align} is stochastically stable at $\delta_{2^n}^{2^n}$. Based on the Lyapunov function for Markovian switching Boolean networks in Subsection \ref{sub2.4}, there exists a Lyapunov function $V_3(z(t),\sigma(t))=\omega_{\sigma(t)}^Tz(t)$ for the closed-loop system (\ref{equ39}) satisfying for $i=1,2,\ldots,r,$
\begin{align*}
&\omega_i=(\omega_{i1}^T,0)^T\in\mathbb{R}^{2^n},~~\omega_{i1}>0\\
&\sum_{j=1}^p\pi_{ij}\tilde{H}_{i,11}^T\omega_{j1}-\omega_{i1}<0,
\end{align*} where $\tilde{H}_{i,11}=[I_{2^n-1}~{\bf 0}_{2^n-1}]H_{i}K\Phi_{2^n}\left[\begin{array}{c}I_{2^n-1}\\{\bf 0}_{1\times( 2^{n}-1)}\end{array}\right]$, $i=1,2,\ldots,r$.

At time $t_k$, to determine the next sampling time $t_{k+1}$ is related to the switching signal $\sigma(t)$ at the sampling time $t_k$. Therefore, the self-triggered scheduling (\ref{equ22}) for Markovian switching BCN (\ref{equ37}) becomes
 \begin{equation}\label{equ44}
\left\{\begin{array}{ll}
u(t)=u(t_k)\in{\cal U}(x(t_k),\sigma(t_k)),&t\in[t_k,t_{k+1}),\\
t_{k+1}=t_k+\tau(x(t_k),\sigma(t_k)),
\end{array}
\right.
\end{equation} where $\tau(x(t_k),\sigma(t_k))$ denotes the time between two consecutive sampling times and ${\cal U}(x(t_k),\sigma(t_k))$ is the possible control set when the state is $x(t_k)$ and the switching signal is $\sigma(t_k)$. Then the self-triggered scheduling (\ref{equ44}) for Markovian switching BCN (\ref{equ37}) can be designed as follows. For $M>0$, if ${\bf E}\{z_{u,t}|z(t_k),\sigma(t_k)\}\neq\delta_{2^n}^{2^n}$ for any $t$ and $u\in\Delta_{2^m}$, denote
\begin{align}
&{\cal U}_M(z(t_k),\sigma(t_k))
=\{u\in\Delta_{2^m}\mid{\bf E}\{V_3\left(z_{u,i}(t_k)\right)|z(t_k),\sigma(t_k)\}\nonumber\\&~~-{\bf E}\{V_3(z_{u,i-1}(t_k))|z(t_k),\sigma(t_k)\}<0, i=1,2,\ldots,M\},
\end{align} where $z_{u,M}(t_k)=(H_{\sigma(t_k+M-1)}u)\cdots(H_{\sigma(t_k)}u)z(t_k)$ and $z_{u,0}(t_k)=z(t_k)$. Otherwise, if there exists some $u\in\Delta_{2^m}$ and a positive integer $N_u\leq M$ such that ${\bf E}\{z_{u,N_u}(t_k)|z(t_k),\sigma(t_k)\}=\delta_{2^n}^{2^n}$, denote
{\small \begin{align}
&{\cal U}_M(z(t_k),\sigma(t_k))
=\{u\in\Delta_{2^m}|\nonumber\\
&{\bf E}\{V_3\left(z_{u,i}(t_k)\right)|z(t_k),\sigma(t_k)\}<{\bf E}\{V_3(z_{u,i-1}(t_k))|z(t_k),\sigma(t_k)\},\nonumber\\
 &{\bf E}\{V_3\left(z_{u,j}(t_k)\right)|z(t_k),\sigma(t_k)\}={\bf E}\{V_3(z_{u,j-1}(t_k))|z(t_k),\sigma(t_k)\},\nonumber\\
 &i=1,2,\ldots,N_u, j=N_u+1,\ldots,M\}.
\end{align}}
 Then $\tau(z(t_k),\sigma(t_k))$ and ${\cal U}(z(t_k),\sigma(t_k))$ are defined as
\begin{align}
&\tau(z(t_k),\sigma(t_k))=\max\{M\mid{\cal U}_M(z(t_k),\sigma(t_k))\neq\emptyset\},\\
&{\cal U}(z(t_k),\sigma(t_k))={\cal U}_{\tau(z(t_k),\sigma(t_k))}(z(t_k),\sigma(t_k)).
\end{align}
\begin{theorem}
Consider Markovian switching BCN (\ref{equ37}). The control strategy in (\ref{equ44}) for (\ref{equ37}) is well defined, i.e., $t_{k+1}>t_k$ for $k=1,2,\ldots.$ Also the system (\ref{equ37}) is stochastically stabilizable at $\delta_{2^n}^{2^n}$.
\end{theorem}

{\em Proof}. Similar to the proof of Theorem \ref{thm1}, we only need to prove for all $z\in\Delta_{2^n}$ and $i\in{\cal R}$, there exists $\bar{u}\in\Delta_{2^m}$ such that ${\cal U}_1(z,i)\neq\emptyset$. Suppose that at some time $t_k$, $z(t_k)=z$. Let $\bar{u}=Kz$, where $K$ is the stabilizing controller given in (\ref{equ38}), then by the properties of Lyapunov function in Subsection \ref{sub2.4} and similar to the proof of Theorem \ref{thm2}, it is easy to get that ${\bf E}\{V_3(z(t_k+1))|z(t_k),\sigma(t_k)\}-V_3(z(t_k),\sigma(t_k))=0$ if $z=\delta_{2^n}^{2^n}$ and ${\bf E}\{V_3(z(t_k+1))|z(t_k),\sigma(t_k)\}-V_3(z(t_k),\sigma(t_k))<0$ if $z\neq\delta_{2^n}^{2^n}$, which implies that $t_{k+1}>t_k$.

Similar to the proof of Theorem \ref{thm2}, the final statement can also be proved.
\hfill$\blacksquare$

\section{conclusion}\label{section5}
In this paper, we studied self-triggered control for three kinds of BCNs, including deterministic, probabilistic and Markovian switching BCNs, in order to deal with the constraint of limited resources. By first reviewing and proposing Lyapunov stability theory for Boolean networks, the self-triggered scheduling was designed based on the decrease of the Lyapunov function between two consecutive samplings and the self-triggered controller was designed, under which the studied BCNs can be ensured to be stabilizable at $\delta_{2^n}^{2^n}$. Some simulation results were presented for illustrating the presented self-triggered strategy.
\bibliographystyle{IEEEtran}
\bibliography{MinMeng}

\begin{thebibliography}{10}
\providecommand{\url}[1]{#1}
\csname url@samestyle\endcsname
\providecommand{\newblock}{\relax}
\providecommand{\bibinfo}[2]{#2}
\providecommand{\BIBentrySTDinterwordspacing}{\spaceskip=0pt\relax}
\providecommand{\BIBentryALTinterwordstretchfactor}{4}
\providecommand{\BIBentryALTinterwordspacing}{\spaceskip=\fontdimen2\font plus
\BIBentryALTinterwordstretchfactor\fontdimen3\font minus
  \fontdimen4\font\relax}
\providecommand{\BIBforeignlanguage}[2]{{%
\expandafter\ifx\csname l@#1\endcsname\relax
\typeout{** WARNING: IEEEtran.bst: No hyphenation pattern has been}%
\typeout{** loaded for the language `#1'. Using the pattern for}%
\typeout{** the default language instead.}%
\else
\language=\csname l@#1\endcsname
\fi
#2}}
\providecommand{\BIBdecl}{\relax}
\BIBdecl

\bibitem{kauffman1969metabolic}
S.~Kauffman, ``Metabolic stability and epigenesis in randomly constructed
  genetic nets,'' \emph{{Journal of Theoretical Biology}}, vol.~22, no.~3, pp.
  437--467, 1969.

\bibitem{kabir2014mathematical}
M.~H. Kabir, M.~R. Hoque, B.~J. Koo, and S.~H. Yang, ``{Mathematical modelling
  of a context-aware system based on Boolean control networks for smart
  home},'' in \emph{{The 18th IEEE International Symposium on Consumer
  Electronics (ISCE 2014)}}, 2014, pp. 1--2.

\bibitem{alexander2003random}
J.~M. Alexander, ``{Random Boolean networks and evolutionary game theory},''
  \emph{{Philosophy of Science}}, vol.~70, no.~5, pp. 1289--1304, 2003.

\bibitem{cheng2014finite}
D.~Cheng, ``On finite potential games,'' \emph{Automatica}, vol.~50, no.~7, pp.
  1793--1801, 2014.

\bibitem{zhang2018incomplete}
X.~Zhang, Y.~Hao, and D.~Cheng, ``Incomplete-profile potential games,''
  \emph{{Journal of The Franklin Institute}}, vol. 355, no.~2, pp. 862--877,
  2018.

\bibitem{Chengd2007}
D.~Cheng and H.~Qi, \emph{{Semi-Tensor Product of Matrices --- Theory and
  Applications}}.\hskip 1em plus 0.5em minus 0.4em\relax Beijing: {Science
  Press}, 2007.

\bibitem{Chengd2011}
D.~Cheng, H.~Qi, and Z.~Li, \emph{{Analysis and Control of Boolean Networks: A
  Semi-Tensor Product Approach}}.\hskip 1em plus 0.5em minus 0.4em\relax
  {Springer}, 2011.

\bibitem{cheng2015receding}
D.~Cheng, Y.~Zhao, and T.~Xu, ``Receding horizon based feedback optimization
  for mix-valued logical networks,'' \emph{{IEEE Transactions on Automatic
  Control}}, vol.~60, no.~12, pp. 3362--3366, 2015.

\bibitem{fornasini2013observability}
E.~Fornasini and M.~E. Valcher, ``{Observability, reconstructibility and state
  observers of Boolean control networks},'' \emph{{IEEE Transactions on
  Automatic Control}}, vol.~58, no.~6, pp. 1390--1401, 2013.

\bibitem{guo2017invariant}
Y.~Guo, Y.~Ding, and D.~Xie, ``{Invariant subset and set stability of Boolean
  networks under arbitrary switching signals},'' \emph{{IEEE Transactions on
  Automatic Control}}, vol.~62, no.~8, pp. 4209--4214, 2017.

\bibitem{liang2017improved}
J.~Liang, H.~Chen, and J.~Lam, ``{An improved criterion for controllability of
  Boolean control networks},'' \emph{{IEEE Transactions on Automatic Control}},
  vol.~62, no.~11, pp. 6012--6018, 2017.

\bibitem{liu2017function}
Y.~Liu, B.~Li, H.~Chen, and J.~Cao, ``{Function perturbations on singular
  Boolean networks},'' \emph{Automatica}, vol.~84, pp. 36--42, 2017.

\bibitem{lu2016pinning}
J.~Lu, J.~Zhong, C.~Huang, and J.~Cao, ``{On pinning controllability of Boolean
  control networks},'' \emph{{IEEE Transactions on Automatic Control}},
  vol.~61, no.~6, pp. 1658--1663, 2016.

\bibitem{toyoda2019mayer}
M.~Toyoda and Y.~Wu, ``{Mayer-type optimal control of probabilistic Boolean
  control network with uncertain selection probabilities},'' \emph{IEEE
  Transactions on Cybernetics}, 2019, doi: 10.1109/TCYB.2019.2954849.

\bibitem{meng2019controllability}
M.~Meng, G.~Xiao, C.~Zhai, and G.~Li, ``{Controllability of Markovian jump
  Boolean control networks},'' \emph{Automatica}, vol. 106, pp. 70--76, 2019.

\bibitem{li2018survey}
H.~Li, G.~Zhao, M.~Meng, and J.~Feng, ``A survey on applications of semi-tensor
  product method in engineering,'' \emph{{Science China Information Sciences}},
  vol.~61, no.~1, p. 010202, 2018.

\bibitem{zhong2019pinning}
J.~Zhong, D.~W. Ho, J.~Lu, and Q.~Jiao, ``{Pinning controllers for activation
  output tracking of Boolean network under one-bit perturbation},'' \emph{IEEE
  Transactions on Cybernetics}, vol.~49, no.~9, pp. 3398--3408, 2019.

\bibitem{li2020perturbation}
H.~Li, X.~Yang, and S.~Wang, ``{Perturbation analysis for finite-time stability
  and stabilization of probabilistic Boolean networks},'' \emph{IEEE
  Transactions on Cybernetics}, 2020, doi: 10.1109/TCYB.2020.3003055.

\bibitem{li2014state}
R.~Li, M.~Yang, and T.~Chu, ``{State feedback stabilization for probabilistic
  Boolean networks},'' \emph{Automatica}, vol.~50, no.~4, pp. 1272--1278, 2014.

\bibitem{li2018set}
F.~Li and L.~Xie, ``{Set stabilization of probabilistic Boolean networks using
  pinning control},'' \emph{{IEEE Transactions on Neural Networks and Learning
  Systems}}, vol.~30, no.~8, pp. 2555--2561, 2019.

\bibitem{meng2017stability}
M.~Meng, L.~Liu, and G.~Feng, ``{Stability and $l_1$ gain analysis of Boolean
  networks with Markovian jump parameters},'' \emph{{IEEE Transactions on
  Automatic Control}}, vol.~62, no.~8, pp. 4222--4228, 2017.

\bibitem{meng2018stability}
M.~Meng, J.~Lam, J.~Feng, and K.~C. Cheung, ``{Stability and stabilization of
  Boolean networks with stochastic delays},'' \emph{{IEEE Transactions on
  Automatic Control}}, vol.~64, no.~2, pp. 790--792, 2019.

\bibitem{huang2020stability}
C.~Huang, J.~Lu, G.~Zhai, J.~Cao, G.~Lu, and M.~Perc, ``{tability and
  stabilization in probability of probabilistic Boolean networks},'' \emph{IEEE
  Transactions on Neural Networks and Learning Systems}, 2020, doi:
  10.1109/TNNLS.2020.2978345.

\bibitem{li2016event}
Q.~Li, B.~Shen, Y.~Liu, and F.~E. Alsaadi, ``{Event-triggered $H_\infty$ state
  estimation for discrete-time stochastic genetic regulatory networks with
  Markovian jumping parameters and time-varying delays},''
  \emph{Neurocomputing}, vol. 174, pp. 912--920, 2016.

\bibitem{yue2017event}
D.~Yue, Z.~Guan, T.~Li, R.~Liao, F.~Liu, and Q.~Lai, ``Event-based cluster
  synchronization of coupled genetic regulatory networks,'' \emph{Physica A:
  Statistical Mechanics and its Applications}, vol. 482, pp. 649--665, 2017.

\bibitem{liu2016sampled}
Y.~Liu, J.~Cao, L.~Sun, and J.~Lu, ``{Sampled-data state feedback stabilization
  of Boolean control networks},'' \emph{{Neural Computation}}, vol.~28, no.~4,
  pp. 778--799, 2016.

\bibitem{zhu2018sampled}
S.~Zhu, Y.~Liu, J.~Lou, J.~Lu, and F.~E. Alsaadi, ``{Sampled-data state
  feedback control for the set stabilization of Boolean control networks},''
  \emph{{IEEE Transactions on Systems, Man, and Cybernetics: Systems}},
  vol.~50, no.~4, pp. 1580--1589, 2020.

\bibitem{liu2019sampled}
Y.~Liu, L.~Tong, J.~Lou, J.~Lu, and J.~Cao, ``{Sampled-data control for the
  synchronization of Boolean control networks},'' \emph{{IEEE Transactions on
  Cybernetics}}, vol.~49, no.~2, pp. 726--732, 2019.

\bibitem{aaarzen1999simple}
K.-E. {\AA}arz{\'e}n, ``{A simple event-based PID controller},'' \emph{IFAC
  Proceedings Volumes}, vol.~32, no.~2, pp. 8687--8692, 1999.

\bibitem{tabuada2007event}
P.~Tabuada, ``Event-triggered real-time scheduling of stabilizing control
  tasks,'' \emph{{IEEE Transactions on Automatic Control}}, vol.~52, no.~9, pp.
  1680--1685, 2007.

\bibitem{lunze2010state}
J.~Lunze and D.~Lehmann, ``A state-feedback approach to event-based control,''
  \emph{Automatica}, vol.~46, no.~1, pp. 211--215, 2010.

\bibitem{lehmann2011event}
D.~Lehmann, \emph{{Event-Based State-Feedback Control}}.\hskip 1em plus 0.5em
  minus 0.4em\relax Logos Verlag Berlin GmbH, 2011.

\bibitem{li2018event_tc}
B.~Li, Y.~Liu, K.~I. Kou, and L.~Yu, ``{Event-triggered control for the
  disturbance decoupling problem of Boolean control networks},'' \emph{{IEEE
  Transactions on Cybernetics}}, vol.~48, no.~9, pp. 2764--2769, 2018.

\bibitem{li2018event}
Y.~Li, H.~Li, and W.~Sun, ``Event-triggered control for robust set
  stabilization of logical control networks,'' \emph{Automatica}, vol.~95, pp.
  556--560, 2018.

\bibitem{zhu2019stabilizing}
Q.~Zhu and W.~Lin, ``{Stabilizing Boolean networks by optimal event-triggered
  feedback control},'' \emph{{Systems \& Control Letters}}, vol. 126, pp.
  40--47, 2019.

\bibitem{yang2019event}
J.~Yang, J.~Lu, L.~Li, Y.~Liu, Z.~Wang, and F.~E. Alsaadi, ``{Event-triggered
  control for the synchronization of Boolean control networks},''
  \emph{{Nonlinear Dynamics}}, vol.~96, no.~2, pp. 1335--1344, 2019.

\bibitem{tong2018robust}
L.~Tong, Y.~Liu, Y.~Li, J.~Lu, Z.~Wang, and F.~E. Alsaadi, ``{Robust control
  invariance of probabilistic Boolean control networks via event-triggered
  control},'' \emph{{IEEE Access}}, vol.~6, pp. 37\,767--37\,774, 2018.

\bibitem{lu2019event}
J.~Lu, J.~Yang, J.~Lou, and J.~Qiu, ``Event-triggered sampled feedback
  synchronization in an array of output-coupled boolean control networks,''
  \emph{IEEE Transactions on Cybernetics}, 2019, doi:
  10.1109/TCYB.2019.2939761.

\bibitem{velasco2003self}
M.~Velasco, J.~Fuertes, and P.~Marti, ``The self triggered task model for
  real-time control systems,'' in \emph{Work-in-Progress Session of the 24th
  IEEE Real-Time Systems Symposium (RTSS03)}, vol. 384, 2003.

\bibitem{wang2009self}
X.~Wang and M.~D. Lemmon, ``{Self-triggered feedback control systems with
  finite-gain ${\cal L}_2$ stability},'' \emph{{IEEE Transactions on Automatic
  Control}}, vol.~54, no.~3, pp. 452--467, 2009.

\bibitem{mazo2010iss}
M.~Mazo~Jr, A.~Anta, and P.~Tabuada, ``{An ISS self-triggered implementation of
  linear controllers},'' \emph{Automatica}, vol.~46, no.~8, pp. 1310--1314,
  2010.

\bibitem{wang2012definition}
Y.~Wang and H.~Li, ``{On definition and construction of Lyapunov functions for
  Boolean networks},'' in \emph{Proceedings of the 10th World Congress on
  Intelligent Control and Automation (WCICA)}, 2012, pp. 1247--1252.

\bibitem{liu2008Hadamard}
S.~Liu and G.~Trenkler, ``{Hadamard, Khatri-Rao, Kronecker and other matrix
  products},'' \emph{{International Journal of Information and Systems
  Sciences}}, vol.~4, no.~1, pp. 160--177, 2008.

\bibitem{li2017lyapunov}
H.~Li and Y.~Wang, ``{Lyapunov-based stability and construction of Lyapunov
  functions for Boolean networks},'' \emph{{SIAM Journal on Control and
  Optimization}}, vol.~55, no.~6, pp. 3437--3457, 2017.

\bibitem{meng20161}
M.~Meng, J.~Lam, J.~Feng, and X.~Li, ``{$l_1$-gain analysis and model reduction
  problem for Boolean control networks},'' \emph{{Information Sciences}}, vol.
  348, pp. 68--83, 2016.

\bibitem{cheng2010realization}
D.~Cheng, Z.~Li, and H.~Qi, ``{Realization of Boolean control networks},''
  \emph{Automatica}, vol.~46, no.~1, pp. 62--69, 2010.

\bibitem{li2019finite}
H.~Li, X.~Xu, and X.~Ding, ``{Finite-time stability analysis of stochastic
  switched Boolean networks with impulsive effect},'' \emph{Applied Mathematics
  and Computation}, vol. 347, pp. 557--565, 2019.

\bibitem{horn2012matrix}
R.~A. Horn and C.~R. Johnson, \emph{{Matrix Analysis}}.\hskip 1em plus 0.5em
  minus 0.4em\relax Cambridge university press, 2012.

\bibitem{Cheng2011}
D.~Cheng, H.~Qi, Z.~Li, and J.~Liu., ``{Stability and stabilization of Boolean
  networks},'' \emph{{International Journal of Robust and Nonlinear Control}},
  vol.~21, no.~2, pp. 134--156, 2011.

\bibitem{li2013state}
R.~Li, M.~Yang, and T.~Chu, ``{State feedback stabilization for Boolean control
  networks},'' \emph{{IEEE Transactions on Automatic Control}}, vol.~58, no.~7,
  pp. 1853--1857, 2013.

\bibitem{wang2019stabilization}
L.~Wang, Y.~Liu, Z.~Wu, J.~Lu, and L.~Yu, ``{Stabilization and finite-time
  stabilization of probabilistic Boolean control networks},'' \emph{{IEEE
  Transactions on Systems, Man, and Cybernetics: Systems}}, 2019, doi:
  10.1109/TSMC.2019.2898880.

\end{thebibliography}

%




\end{document}